\documentclass[pre,aps,pacs,twocolumn]{revtex4}
\usepackage{epsfig}
\usepackage{bm}
\usepackage{color}
\newcommand{\ud}{\mathrm{d}}
\newcommand{\B}[1]{{\bm{#1}}}

\usepackage[latin1]{inputenc}
\newcommand{\beq}{\begin{equation}}
\newcommand{\eeq}{\end{equation}}
\newcommand{\bea}{\begin{eqnarray}}
\newcommand{\eea}{\end{eqnarray}}

\begin{document}
\title{Mechanical Properties of Glass Forming Systems}
\author{Valery Ilyin, Nataliya Makedonska, Itamar Procaccia and Nurith Schupper }
\affiliation{Department of Chemical Physics, The Weizmann
Institute of Science, Rehovot 76100, Israel }
\begin{abstract}
We address the interesting temperature range of a glass forming system where the mechanical
properties are intermediate between those of a liquid and a solid. We employ an efficient Monte-Carlo method to calculate the elastic moduli, and show that in this range of temperatures the moduli are finite for short times and vanish for long times, where `short' and `long' depend on the temperature. By invoking some exact results from statistical mechanics we offer an alternative method
to compute shear moduli using Molecular Dynamics simulations, and compare those to the Monte-Carlo method. The final conclusion is that these systems are not ``viscous fluids" in the usual sense, as  their actual time-dependence concatenates solid-like materials with varying local shear moduli. 
  \end{abstract}
\maketitle

Traditionally the term ``glass-transition" refers to the spectacular increase in viscosity when
the temperature $T$ of a liquid is lowered within a relatively narrow temperature range. Some
authors even proposed to call an amorphous medium a ``glass" when its viscosity increases
above $10^{12}$ Pa s \cite{06Dyre}. For sufficiently high temperatures such glass-former are
liquids with a finite viscosity. For sufficiently low temperatures these systems are solids, although they fail to exhibit any crystalline order. Here we address the range of temperature in between, and attempt to clarify the state of matter that is observed there. Over the years a qualitative picture seems to have emerged: already Goldstein in 1969 \cite{69Gol} proposed that in the intermediate temperature range the potential energy surface exhibits various minima separated by relatively high potential barrier (with respect to $k_B T$). Stillinger and Weber expounded this further, discussing the existence of ``inherent states". The dynamics is trapped for a while in an
inherent state and then hops, via a saddle, to another ``inherent state" \cite{82SW}. Demonstrations of such events were presented for example in \cite{00SSDG}. The aim of this Letter is to explore the
nature of these ``inherent states" from the point of view of their mechanical properties. We will offer
a complementary point of view focusing on the elastic moduli
of the material. An accepted notion of a liquid is a medium that cannot support shear, and thus
by definition has a zero shear modulus. In this Letter we propose that it is advantageous to 
study the shear moduli in glass formers, since the notion of viscosity loses its usual meaning as a local
descriptor of the state of flow in this range of temperatures, as is explained below. In contradistinction, the shear moduli lend themselves to accurate measurement also locally in time throughout the temperature range. Moreover, the typical jamming that occurs
at low temperatures can be overcome when computing the shear moduli by employing an efficient
Monte Carlo method \cite{03WTK} as is described below. Using this point of view we will be able
to present in this Letter accurate calculations showing that there exists an interesting temperature
range where the elastic moduli are finite when computed for short times, but tending to zero when computed for longer and longer times. The appropriate length of time should be quantified with respect to the statistics of the (relatively rare) relaxation events in which the system jumps from one state to the other. We reiterate that viscosity in the local sense does not exist. 

For concreteness and in order to be able to check our calculations with respect to others we chose to work here on the well studied glass former obtained from a two-dimensional binary mixture of discs interacting via a soft $1/r^{12}$ repulsion with a ``diameter" ratio of 1.4 \cite{89DAY,99PH}. The particles have the same mass $m$, but half of the particles are
`large' with `diameter' $\sigma_2=1.4$ and half of the particles are `small' , with `diameter' $\sigma_1=1$. The three pairwise additive interactions are purely repulsive:
\begin{equation}
u_{ab} =\epsilon \left(\frac{\sigma_{ab}}{r}\right)^{12} \ , \quad a,b=1,2 \ , \label{potential}
\end{equation}
where $\sigma_{aa}=\sigma_a$ and $\sigma_{ab}= (\sigma_a+\sigma_b)/2$. The cutoff radii of 
the interaction are set at $4.5\sigma_{ab}$. The units of mass, length, time and temperature are $m$, $\sigma_1$, $\tau=\sigma_1\sqrt{m/\epsilon}$ and $\epsilon/k_B$, respectively, with $k_B$ being
Boltzmann's constant. 
Ref. \cite{99PH} found, using molecular dynamics simulations in the isothermal-isobaric ($NPT$) ensemble, that for temperature $T>0.5$ the system is liquid and  for lower temperatures dynamical relaxation slows down. A precise glass transition had not been identified in \cite{99PH}. In \cite{06ABIMPS,07HIMPS} it
was argued on the basis of statistical mechanics that there exists a typical length scale that
grows exponentially fast when the temperature decreases. Associated with this fastly growing scale there exists a fastly growing relaxation time, such that below a certain temperature the system is jammed for all practical purposes. Here we will shed further light on this phenomenon.

To measure elastic moduli at any temperature one employs the relation between the strain fluctuations and the elastic properties of a system.  
In the frame of the isothermal-isobaric ensemble the strain simultaneous correlation functions are  given by:
\beq
<\epsilon_{ij}\epsilon_{kl}>=\frac{\int \epsilon_{ij}\epsilon_{kl}
\exp(-\frac{\Delta G}{ T})\ud\epsilon_{xx}\ud\epsilon_{yy}
\ud\epsilon_{xy}}{Z},  \label{corr}
\eeq
where $Z$ is the following partition function:
\beq
Z=\int
\exp(-\frac{\Delta G}{ T})\ud\epsilon_{xx}\ud\epsilon_{yy}
\ud\epsilon_{xy} \,  \label{partition}
\eeq
and $\epsilon_{ij}$ are the components of the strain tensor. $\Delta G$ is the difference of Gibbs free energy between the strained and the reference state. 
In the linear elastic approximation for a 2-dimensional system it takes on the form  \cite{03WTK}:
\begin{equation}
\frac{\Delta G }{A_0}=\frac{1}{2}\left[\lambda(\epsilon_{xx}+\epsilon_{yy})^{2}+\mu'(\epsilon_{xx}-\epsilon_{yy})^2+4\mu \epsilon_{xy}^2\right],  \label{DeltaG} 
\end{equation}
where $\lambda$, $\mu$ and $\mu'$ are the elastic moduli,  and $A_{0}$ is the area of the system in the reference state with respect  to which the free energy is computed. With this form of the free energy
Eq. (\ref{corr}) reduces to simple Gaussian integrals yielding the wanted relations \cite{58LL}, \cite{82PR}. Here we are interested mostly in the shear modulus $\mu$ which is defined by:
\begin{equation}
\mu=\frac{1}{4<\epsilon_{xy}^{2}>}\frac{ T}{A_{0}} \ ,
\label{shearmod}
\end{equation}

The strain fluctuation in (\ref{shearmod}) can be evaluated using numerical simulations.
An efficient Monte Carlo method aimed at calculating this quantity in $NPT$ ensemble  was described in \cite{03WTK}. In essence, the method introduces strains into the simulation box by first defining a square box of unit area where the particles are at positions $\B s_i$. Then one defines a linear transformation $\B h$, taking the particles to positions $\B r_i$ via $\B r_i =\B h \B s_i$. The  area of the system becomes the determinant $|\B h|$. In order to prevent rotations of the 
simulation box, the matrix $h$ should be symmetric. After the application of the transformation
one performs $n$ standard Monte-Carlo moves, after which the transformation $\B h$ changes according to 
\beq
h_{ij}^{new}=h_{ij}^{old}+\Delta h_{max}(2\xi_{ij}-1) \ , \label{hnew}
\eeq
where $\Delta h_{max}$ is the maximum allowed change of a matrix 
element and $\xi_{ij}$ is a random number uniformly distributed between 
0 and 1. At this point one performs additional $n$ standard Monte Carlo steps etc. 
The strain tensor in this procedure is calculated by \cite{r4}:
\beq
{\bf \epsilon}=\frac{1}{2}({\bf h}_{0}^{ -1}
{\B h}\B h {\bf h}_{0}^{-1}-{\bf I}) \ , \label{defeps}
\eeq
where ${\bf h}_{0}=<{\bf h}>$ is the reference box matrix which is obtained from
averaging $\B h$ until the time of evaluation, and ${\bf I}$ is the unit matrix. In our calculation we chose $n=100$ for the number
of Monte-Carlo steps between strain transformations. Note that the strain $\B \epsilon$ depends on
time both due to $\B h$ itself and due to (the possibly slowly convergent) time dependent $\B h_0$.
\begin{figure}
\centering
\epsfig{width=.49\textwidth,file=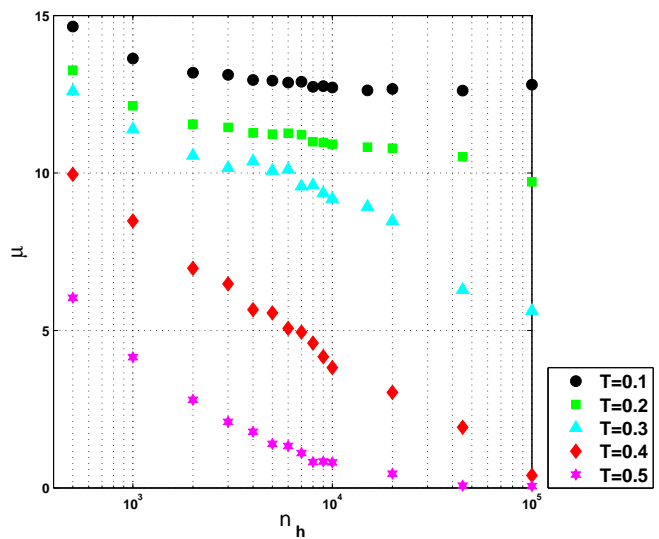}
\caption{(Color online). The shear modulus computed from Monte-Carlo simulations as a function of the 
number of $\B n_{h}$ sweeps, for different values of the temperature}
\label{shear}
\end{figure}

In Fig. \ref{shear} we present the results of our calculations for the shear modulus $\mu$. We have
used 1024 particles in the 2-dimensional simulation cell.
After an initial run of 5000 $\B n_{h}$ sweeps (undisplayed in Fig. \ref{shear}),
these quantities were measured as a function of the number of times $n_{\B h}$
of $\B h$ sweeps, for the five temperatures $T=0.1$, 0.2. 0.3, 0.4 and 0.5 and
the pressure $P=13.5$ (in units of $\epsilon / \sigma_{1}^{2}$). Clearly, for T=0.1 the shear modulus converges to a time-independent value, indicating the existence of a real solid. For T=0.4 and T=0,5 the shear modulus converges to
zero, indicating the existence of a fluid. For T=0.2 and 0.3 the shear modulus did not converge, continuing to decrease as a function of $n_{\B h}$, possibly approaching zero asymptotically, but
maybe stabilizing at a finite value after a very long time.

\begin{figure}
\centering
\epsfig{width=.40\textwidth,file=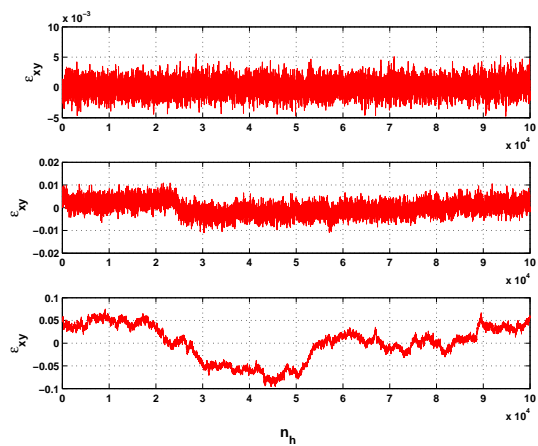}
\caption{(Color online). Trajectories of measured values of  $\epsilon_{xy}$ in the Monte-Carlo simulations as a function of  $n_{\B h}$. Upper panel: $T=0.1$. Middle panel: $T=0.3$. Lower panel: $T=0.5$.}
\label{MC}
\end{figure}

In Fig. \ref{MC} we plot the actual evolution of $\epsilon_{xy}$ as a function of $n_{\B h}$. For $T=0.1$ these values fluctuate around a constant value, but for $T=0.3$ we note the sharp transition between
different ``constant" values that now fluctuate in time. The trajectory at $T=0.5$ represents the almost
free fluctuations in the value of $\epsilon_{xy}$ leading to a huge value of $\langle \epsilon^2_{xy}\rangle$ which is associated with a vanishingly small shear modulus.

\begin{figure}
\centering
\epsfig{width=.40\textwidth,file=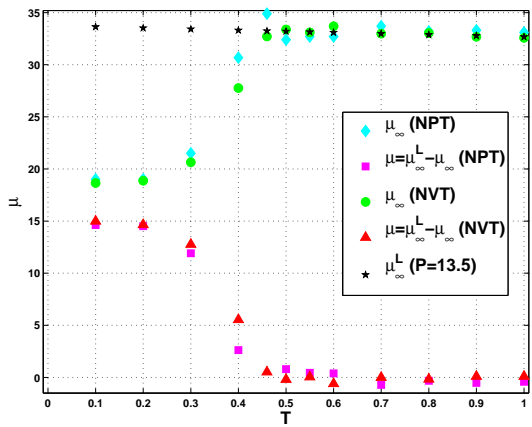}
\caption{(Color online). Results of Molecular Dynamics Simulations. Black stars: exact calculation of
$\mu^L_{\infty}$. Cyan diamonds and green circles: simulation results for
$\mu_\infty$ . Magenta squares and red triangles: the shear modulus
according to Eq. (\ref{diff})}. 
\label{MD}
\end{figure}
\begin{figure}
\centering
\epsfig{width=.37\textwidth,file=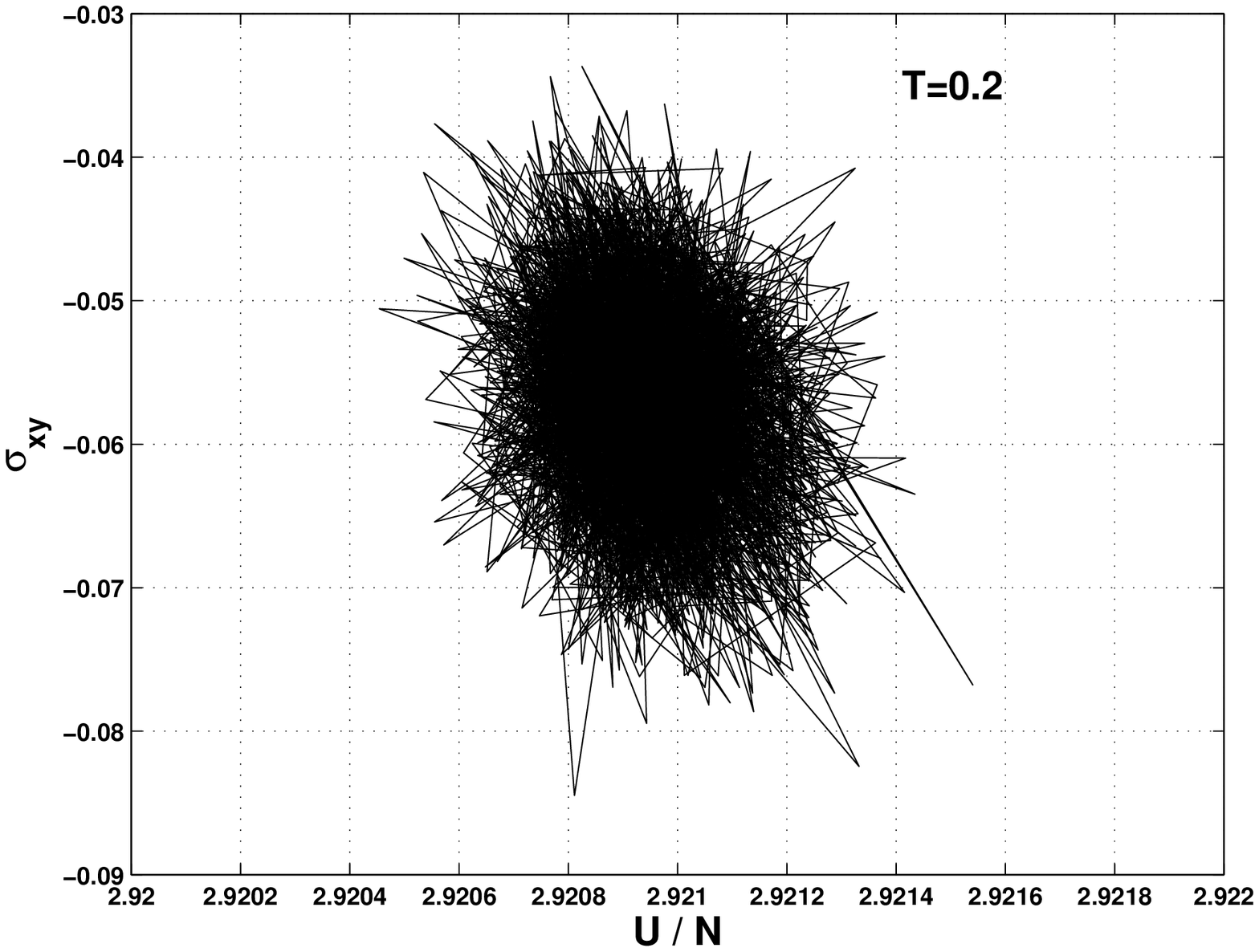}
\epsfig{width=.37\textwidth,file=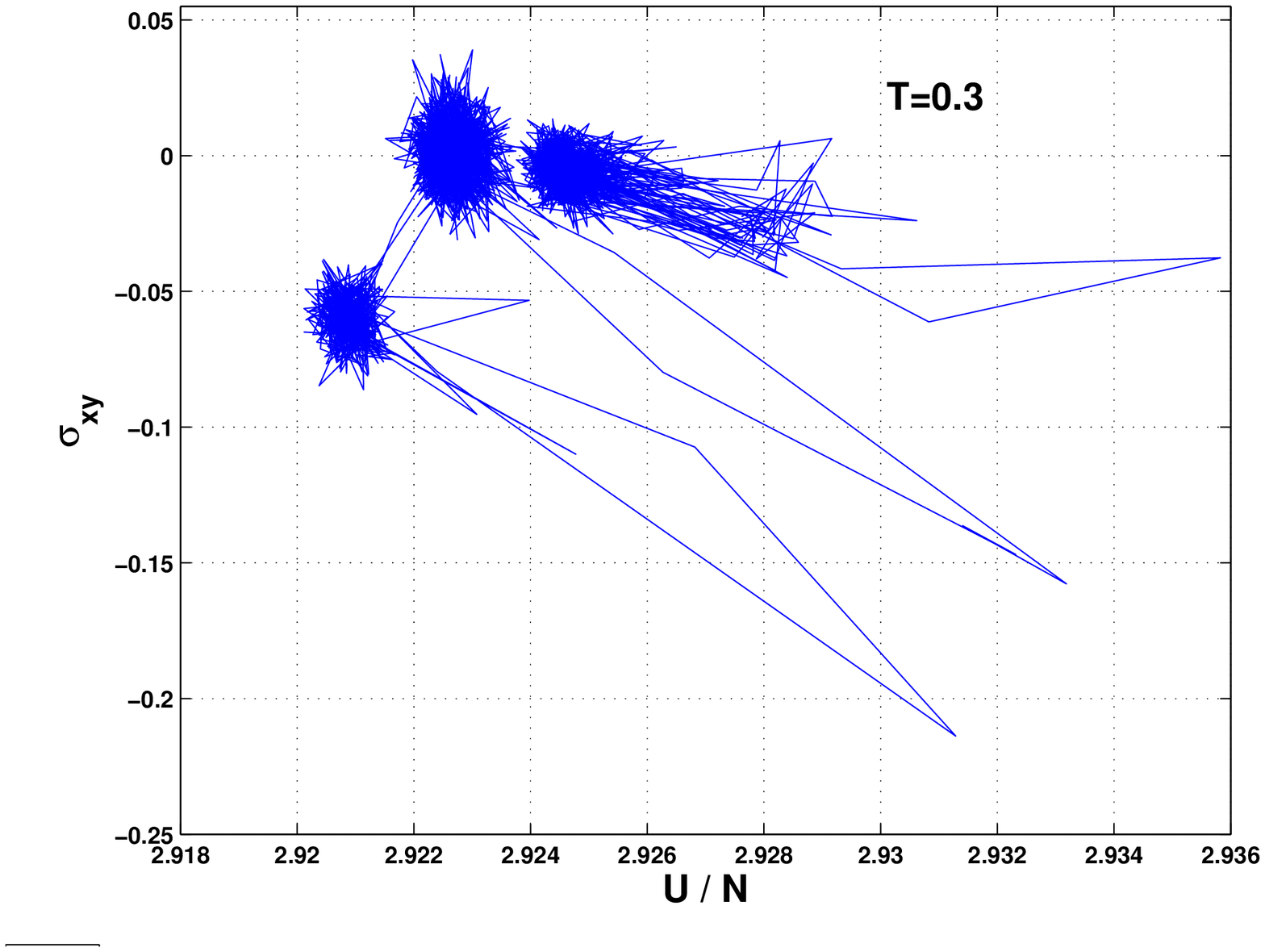}
\epsfig{width=.37\textwidth,file=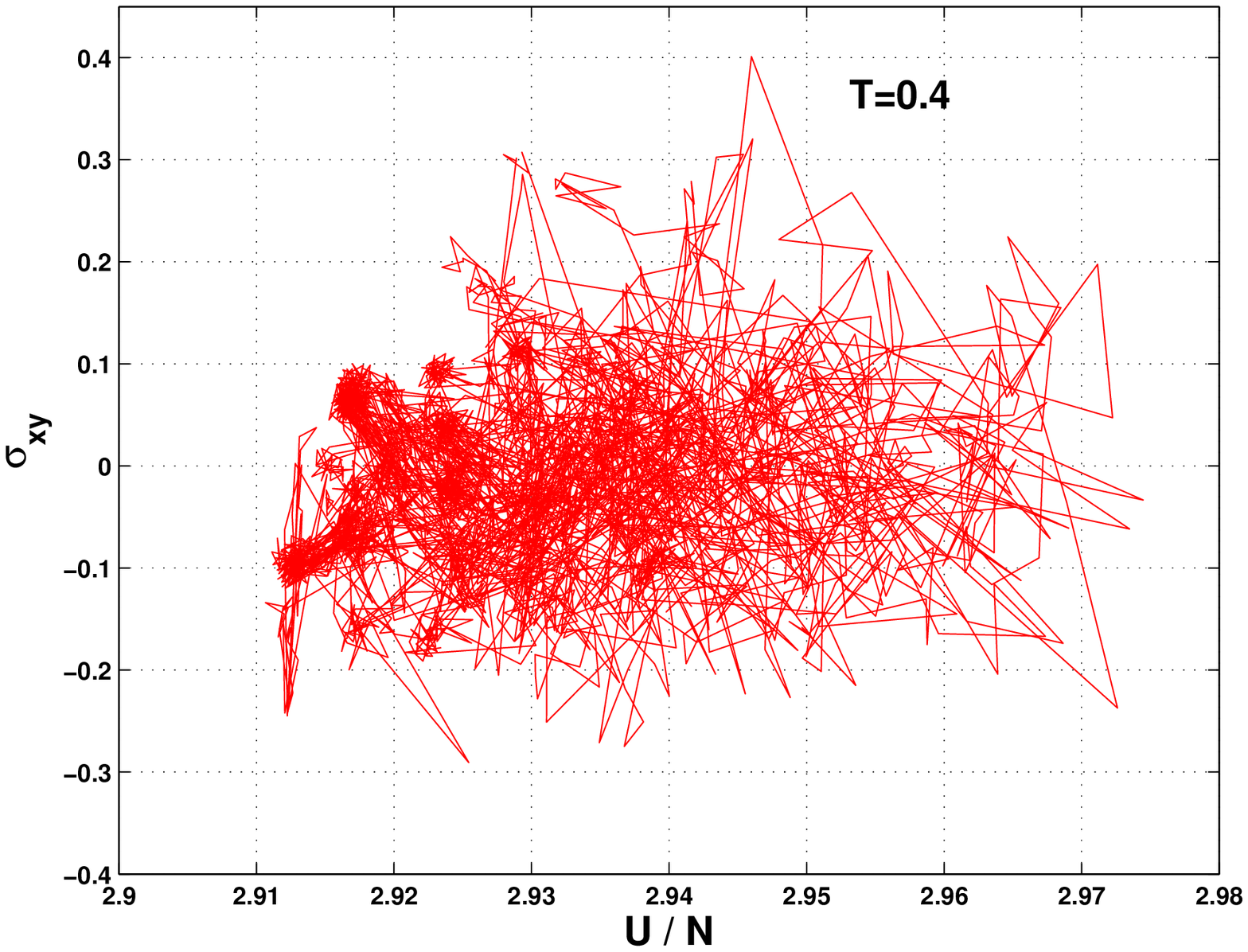}
\caption{(Color online). Trajectories from Molecular Dynamics simulations in
  the stress-energy plane, for three different temperatures. $ U/N$ is
  the potential energy of the system per particle. Each point represents an average over a duration of 50$\tau$, connected with straight lines. The time of simulation  is $1.5\times 10^5\tau$. During this time with T=0.2 the system fluctuates around a single ``solid state". During the same simulation time when T=0.3 the system hops between three different and distinct ``solid states". When T=0.4 the fluctuations are so large that the shear modulus averages to zero during the simulation time. } 
\label{jumps}
\end{figure}

To complement these findings we present now a different method to compute shear moduli.  To this aim we invoke  some
exact results from statistical mechanics. Two important results are due to Zwanzig \cite{65ZM} and to Wallace \cite{00Wal}, relating the shear modulus to other computable objects. The first object is
the so-called ``infinite frequency shear modulus" $\mu_\infty$ which is defined as

\beq
\mu_{\infty}\equiv \frac{A_{0}}{ T}<\sigma_{xy}^2>, \label{Ginfty}
\eeq
where the microscopic shear stress is given by:
\beq
\sigma_{xy}(t)=\frac{1}{A_{0}}\cdot (\sum\limits_{i}\frac{p_{i,x}\cdot p_{i,y}}
{m_{i}}+\sum\limits_{i} r_{i,x}\cdot F_{i,y}) \ , \label{stress}
\eeq
where $p_{i,\alpha}$ is the $\alpha$ component of the momentum of particle $i$ 
and $F_{i,\alpha}$ is the $\alpha$ component of the force exerted on 
particle $i$ \cite{87AT}.  Zwanzig had considered a different quantity, which is referred to by the same name "infinite frequency shear modulus" but which is only appropriate for liquids,  denoted here $\mu_\infty^L$. In two dimensions this object reads 
\beq
\mu^L_{\infty}\equiv \rho T+\frac{1}{8 A_{0}}
<\sum\limits_{i<j}[\frac{1}{r} \frac{\partial}{\partial r}(r^3
\frac{\partial \phi(r)}{\partial r})]_{ij}> \ . \label{GL}
\eeq
For  interparticle potentials of the form $1/r^n$ (n=12 in (\ref{potential})) the expression (\ref{GL}) yields:
\beq
\mu^L_{\infty}=\rho\ T+\frac{n-2}{4} (P - \rho T) \ . \label{ourGL}
\eeq
Wallace obtained in 2000 the remarkable result (apparently uncited even once)
\cite{00Wal}:
\beq
\mu=\mu_{\infty}^{L}-\mu_{\infty} \ , \label{diff}
\eeq 
where the shear modulus $\mu$ is the same as defined by (\ref{shearmod}).
We thus learn that statistical mechanics provides us an indirect way to measure the shear modulus as the difference between the exactly calculable (\ref{ourGL})
(in our case with $n=12$) and the simulationally accessible $\mu_\infty$ Eq. (\ref{Ginfty}).

We performed Molecular Dynamics simulations in the canonical (NVT) ensemble with $N=256$ particles in a square simulation box.  The equations of motion were integrated using a third order Gear predictor-corrector algorithm, a constant temperature was preserved using a velocity rescaling method \cite{87AT}, at each temperature the density was chosen in accordance with the simulations results in NPT ensemble as described in \cite{99PH}. We also followed this reference in correcting the shear modulus for the frozen-in stress. Our results  and the results obtained in \cite{99PH} are presented in Fig. \ref{MD}.
We see that as long as the system is liquid ($T>0.4$), the Zwanzig calculation agrees exactly with
our simulation of $\mu_\infty$, predicting zero shear modulus, as expected. For smaller values of $T$,
$\mu_\infty$ deviates from the liquid value, and the difference is the Molecular Dynamics estimate
for the shear modulus. As before, in the interesting region of temperatures this method indicates a finite value of the shear modulus, but this again depends on the simulation time, and still needs to be interpreted. 

To understand what is the state of matter that is behind these results we present in Fig. \ref{jumps}
the actual trajectory of the Molecular dynamics simulations in the stress-energy plane for three
different temperatures and for the same simulation time. We see that at $T=0.2$ the system fluctuates
around one distinct state, giving us a non-zero value of $\mu_\infty$ as computed from (\ref{Ginfty}).
For all that one can judge we have a solid with a finite value of its shear modulus. 
Changing the temperature to $T=0.3$ we recognize that the system fluctuates around one such
``solid-like" state, but then jumps to another such ``solid-like" state, and then another such state; for this time of the simulation one resolves at $T=0.3$ three states each of which would have given us a sharply defined shear modulus for the time of life of that state. Upon averaging over longer periods we get contributions from different ``solid-like'' states and the shear modulus becomes dependent on the 
time of averaging, as seen in Figs. \ref{shear} and \ref{jumps}. For $T=0.4$ the trajectory now fills up
an extended region in the stress-energy plane, and one can see why the shear modulus must average to zero for longer averaging times (see also Fig. \ref{MC}). Anyway we look at the dynamics the emerging picture is the same: between real solid and real liquid
the system is locally not a ``viscous fluid". Rather, the trajectory concatenates relatively long period of times where the system behaves like a solid, interconnected by relatively short period of times where
the system flows between these states. Clearly, a viscous fluid behaves very differently, responding
to stress by a viscous flow, be it as sluggish as one wishes. Here, most of the time, the system does not
respond to stress, except in the narrow corridors of flow which become rare when the temperature goes down and more common when the temperature warms up. Of course this does not mean that in the sense of long time averaging the notion of viscosity cannot be resurrected, but locally in time this
is impossible.

The question that remains is how to describe macroscopically this state of matter. In our opinion
a promising approach was described in \cite{02GCGP}, where all the accessible stationary points of the potential energy surfaces ${\cal V}(\B r_1\cdots \B r_N)$ for all the $N$ particles were obtained by solving the $3N$ nonlinear equations $\partial {\cal V}/\partial \B r_i=0$. Having a complete enumeration of the accessible minima, the saddles and the eigenvalues of the Hessain matrix at the saddle (cf.  \cite{02GCGP} for details) one can hope to analyze the relative length of time that the system might stay in each of its ``solid-like" minima. In that case one can attempt to answer the question whether a long time average would yield a finite or a zero shear modulus or at any rate describe the mechanical behavior of the system for a finite time. Such a program is becoming more and more realistic with the increase in computer power, and appears unavoidable if one wants to make real progress in understanding the interesting range of temperatures between liquid and solid in amorphous glass-formers.

\end{document}